\documentclass[english,keywords,amsmath,amssymb,showpacs]{revtex4}
\usepackage[T1]{fontenc}
\usepackage[latin1]{inputenc}
\usepackage{babel}%
\usepackage{graphicx}
\usepackage{color}
\usepackage{bm}
\usepackage{longtable}
\usepackage{amsmath}
\usepackage{amsfonts}
\usepackage{amssymb}

\begin{document}
\title{Faithful Pointer for qubit measurement}
\author{Asmita Kumari and A. K. Pan}

\affiliation{National Institute Technology Patna, Ashok Rajhpath, Bihar 800005, India}

\begin{abstract}
In the context of von Neumann projective measurement scenario for a qubit system, it is widely believed that the mutual orthogonality between the post-interaction pointer states is the sufficient condition for achieving the ideal measurement situation. However, for experimentally verifying the observable probabilities, the real space distinction between the pointer distributions corresponding to post-interaction pointer states play crucial role. It is implicitly assumed that mutual orthogonality ensures the support between the post-interaction pointer distributions to be disjoint. We point out that mutual orthogonality (formal idealness) does \emph{not} necessarily imply the real space distinguishability  (operational idealness), but converse is true. In fact, for the commonly referred Gaussian wavefunction, it is possible to obtain a measurement situation which is formally ideal but fully nonideal operationally. In this paper, we derive a class of pointer states, that we call faithful pointers, for which the degree of formal (non)idealness is equal to the operational (non)idealness. In other words, for the faithful pointers, if a measurement situation is formally ideal then it is operationally ideal and vice versa.
\end{abstract}
\maketitle
\section{Introduction}
Measurement in quantum mechanics(QM) is in real contrast to its counterpart in classical world. This is due to the fact that any measurement described by QM entails an interaction between the measuring apparatus and the observed system resulting in the state of the measured system to be necessarily entangled with the state of the observing appratus. In early days of the developement of QM, there was an intense debate whether the apparatus needs to be  treated quantum mechanically. However, there is no logical reason why an appratus should be treated as classical object unless a 'cut' is available between classical and quantum within the formalism of QM. The quantum mechanical modelling for measurement process was first put forward by von Neumann\cite{von} in his book where the measuring device was first treated quantum mechanically.

For simplicity, in this paper we consider von Neumann projective measurement scenario for qubit system and restrict our discussion for the case of pure state. The argument presented here can easily be generalised for the mixed state but the essence of it can be fully understood for the case of pure state. The essential theory of quantum measurement of a qubit system is as follows. Let the initial state of a system be given by $|\chi(t=0)\rangle=\alpha\left|\uparrow\rangle_{x}+\beta\right|\downarrow\rangle_{x}$, where $|\uparrow\rangle_{x}$ and $|\downarrow\rangle_{x}$ are the eigenstates of ${\sigma}_{x}$ with $\alpha$ and $\beta$ satisfying the relation $|\alpha|^{2}+|\beta|^{2}=1$. The total system-apparatus wave function can be written as $\Psi\left(x,t=0\right)=\psi_{0}\left(x\right)\otimes|\chi(t=0)\rangle$, where the $\psi_{0}({x})$ is the spatial part, served the purpose of apparatus. Now, for measuring the system observable $\hat{\sigma_{x}}$, a suitable Hamiltonian is taken to be $\mathcal{H}=g(t)\hat{x}\otimes\hat{\sigma_{x}}$, where $g(t)$ is the coupling constant with $\int_{0}^{t}g(t)dt=g$ and $\hat{x}$ is the apparatus observable. After measurement interaction, the final state is entangled between system and apparatus, can be written as  
 
\begin{equation}
\Psi({x},t)=\alpha\psi_{+}(x,t)\otimes\left|\uparrow\right\rangle_{x}+\beta\psi_{-}(x, t)\otimes\left|\downarrow\right\rangle_{x}
\end{equation}     
where $\psi_{+}(x,t)$ and $\psi_{-}(x,t)$ are the apparatus states after the interaction at time $t$.\\

Now, within von Neumann projective measurement scenario it is commonly believed that a measurement can be considered `ideal(strong)' if the post-interaction apparatus states  $\psi_{+}({x},t)$ and  $\psi_{-}({x},t)$ are \emph{mutually orthogonal}. Such condition formally establishes the perfect one-to-one correspondence between the system and apparatus, so that, the value of the system can unambiguously be inferred from the post-interaction pointer states. Mathematically, one can define the relevant quality in terms of the modulous of the inner product between $\psi_{+}({x},t)$ and  $\psi_{-}({x},t)$ is given by
\begin{equation}
|I|=|\int_{-\infty}^{+\infty}\psi^{*}_{+}({x},t) \psi_{-}({x},t)d{x}|
\label{innerprod}
\end{equation}\\
The value of $|I|$ ranges from $0$ to $1$ and provides the formal (non)idealness of the measurement. Clearly, $|I|\approx 0$ and $|I|\approx 1$ imply the measurement situation is formally \textit{ideal} and \textit{fully nonideal} respectively.

However, there is an implicit assumption in von Neumann projective measurement scenario is that when $|I|\approx 0$ the associated distributions corresponding to the $\psi_{+}({x},t)$ and  $\psi_{-}({x},t)$ are expected to be disjoint in the real space. Such a requirement can be termed as \textit{operational idealness}, which can be defined as  the overlap between $|\psi_{+}({x},t)|^{2}$ and $|\psi_{-}({x},t)|^{2}$ is given by
\begin{equation}
M=\int_{-\infty}^{+\infty}\sqrt{|\psi_{+}({x},t)|^{2}|\psi_{-}({x},t)|^{2}}d{x} 
\end{equation}
The value of $M$ ranges from $0$ to $1$, and $M\approx 0$ implies that associated real space pointer distributions have disjoint support.\\

Now, since for any two complex functions $f_1(z)$ and $f_2(z)$, the identity $\int_{-\infty}^{\infty}|f_1(z)||f_2(z)|dz \geq |\int_{-\infty}^{\infty}f_1^{\ast}(z) f_2(z)dz|$ always holds, then from Eqs.(2-3), we can say that the inequality $M\geq |I|$ is always satisfied. This implies that a particular measurement situation can be formally ideal but may \emph{not} be operationally ideal. In other words, the post-interaction apparatus states can be separable in complex space but their associated distributions may \textit{not} be distinguishable in real space. However, it seems to be a natural demand that in the actual experiments, the real space distinctness is the key criterion for verifying the result of the measurements. Hence, the operationally ideal measurement situation is \emph{stronger requirement} than the formally ideal one. Interestingly, it can even be possible to show that for a particular choices of initial apparatus state and coupling strength, the value of $M\approx 1$ while $|I|\approx 0$, i.e., even if post-interaction system-apparatus is perfectly entangled($|I|\approx 0$), the corresponding distributions $|\psi_{+}({x},t)|^{2}$ and $|\psi_{-}({x},t)|^{2}$ may \emph{not} be separated in real space. In such a case, the use of projection postulate would provide wrong formula of observable probabilities. 

It may then be interesting to see how  the empirically verifiable observable probabilities can be calculated in a nonideal measurement. For the case of the measurement of $\hat{\sigma_{x}}$ by using a suitable Stern-Gerlach (SG)  setup \cite{scully,platt,home,pan,hsu}and  by keeping in mind the Gaussian wave packet as the initial pointer having zero mean position, one can define a quantity  called error measure,
\begin{eqnarray}
E=\int_{x=0}^{+\infty}|\psi_{-}({x},t)|^{2}dx =\int_{x=-\infty}^{0}|\psi_{+}({x},t)|^{2}dx 
\end{eqnarray}
whose value ranges from $0$ to $1/2$ corresponding to \emph{ideal} situation and to fully \emph{nonideal} situation respectively. The error measure $(E)$ is intuitively related to the real space overlap ($M$). When $M\approx 0$ and $M\approx 1$ the values of $E$ are expected to be $0$ and $1/2$ respectively \cite{home}.

Using the error measure given by Eq.(4), one obtains the probability of finding the particles with $\left|\uparrow\right\rangle_{x}$  in the \emph{upper channel} (say, $p^u_{+x}$) of the SG setup and probability of finding the particle with $\left|\downarrow\right\rangle_{x}$ in the \emph{lower channel} (say $p^d_{-x}$) in any given nonideal situation are respectively given by
\begin{eqnarray}
p^u_{+x}=(1-E)|\alpha|^{2} \   \  ;  \   \ p^d_{-x}=(1-E)|\beta|^{2}
\label{nonideal}
\end{eqnarray}
which clearly reduces to the ideal projective case when $E=0$. The above results can also be obtained by suitably defining the POVMs by considering the concept of generalized measurement. By using the familiar notion of unsharp measurements \cite{ozawa,busch} one can derive the relevant POVMs are given by $\Pi_{\pm}=(\mathbb I-(1-2E)P_{\pm x})/2$, where $P_{\pm x}$ are projectors corresponding to $\sigma_x$ measurements. So  that the probability of getting the particle in upper and lower channel are $p_{+}=Tr[|\chi(t=0)\rangle\langle\chi(t=0)|\Pi_{+}]= (1-E)|\alpha|^2 + E |\beta|^2$ and $p_{-}=Tr[|\chi(t=0)\rangle\langle\chi(t=0)|\Pi_{-}]= (1-E)|\beta|^2 + E |\alpha|^2$ respectively. For operationaly ideal measurement ($E=0$), $p_{+}=p^u_{+x}=|\alpha^2|$ and $p_{-}=p^d_{-x}=|\beta^2|$, as expected.\\ 

However, if we restrict ourselves in the von Neumann projective measurement scenario, the requirement of $E$ (or $M$) for calculating observable probabilities can be viewed as a kind of incompleteness of the von Neumann measurement formalism by requiring an error measure to obtain the aforementioned probabilities. Such an error measure is brought from the outside the formalism of von Neumann measurement scenario. It would then be interesting to find a class of pointer for which formally ideal measurement situation ensures the operational ideality of the measurement. In this paper, we derive such a class of pointers for which $M=|I|$ is always satisfied.\\

In the context of non-ideal measurement of a qubit, in a recent work \cite{silva}, the family of optimal pointers is derived for which a suitably defined trade-off relation between the information and disturbance created by the measurement is satisfied. It is shown that such a trade-off relation cannnot be satisfied by tunning coupling strength only, a special class of optimal pointers is required for that purpose. They showed that the square and the Gaussian pointers are not useful for that purpose. A family of optimal pointers satisfying their trade-off relation derived \cite{silva}.\\

 On the other hand, motivation of our work is to derive a class of faithful pointer for which formal idealness ensures the operational idealness. For such a class of pointers, we do not need to define POVMs because the measurement situation is always operationaly ideal if it is formaly ideal. We also note here that the Gaussian class of pointer do not serve as a faithful one. This is due to fact that a for choice of relevant parameter(such as, interaction time, field strength, etc.), the value of $M$ can be made \textit{unity}, while $\textit{I}$ can be made \textit{zero}. Hence, if for a class of pointers, the relation $M=|I|$ is satisfied for any arbitrary coupling, then the operational (non)idealness is equal to formal one. In that case the mutual orthogonality is sufficient for ensuring the measurement situation to be ideal and one can use the projective postulate to calculate the observable probabilities. Such a class of pointers can be considered as a faithful one for von Neumann projective measurement scenario. Moreover, for those pointers, one does not need to consider an operational quantity(such as, $M$ or $E$) from outside the von Naumann formalism for obtaining the experimentally verifiable probabilities.

 The paper is organized as follows. In Section II, we show that the Gaussian-class of pointers cannot be served as a faithful pointer. In Section III, we derive a class of faithful pointers and for which it is shown that if the measurement situation is formally ideal then it is operationally ideal. We conclude in Section IV by providing a brief summary.

\section{The case of Gaussian and squeezed pointers}

The Gaussian pointers are commonly used as an pointer state for describing quantum measurement theory \cite{home,pan,bohm}. We show that such states cannot be served as a suitable pointer. For this, let $\psi_{0}(x)$ is the initial apparatus state represented by Gaussian wave packet which is peaked at $\textit{x}=0$ at $t=0$.
\begin{equation}
\psi_{0}(x)=(2 \pi \sigma^{2}_{0})^{-\frac{1}{4}}\exp{(-{\textit{x}^{2}}/4 \sigma^{2}_{0})}
\end{equation}
where $\sigma_0$ is the initial width of the associated wave packet.\\ 

If the interaction Hamiltonian is taken to be $\textit{H}= g \widehat{x} \otimes \widehat{\sigma_{x}}$, the post-interaction apparatus states $\psi_{+}({x},t)$ and $\psi_{-}({x},t)$ given by Eq.(1) at time $t$ are of the form
\begin{equation}
\psi_{\pm}({x},t)=(2 \pi s^{2}_{t})^{-\frac{1}{4}}\exp{[-\frac{(x \mp \frac{g t}{2})^{2}}{4 \sigma_{0} s_{t}} \pm i k_{x} x]}
\end{equation}

where   $s_{t}=\sigma_{0}(1+ i t/2  \sigma^{2}_{0})$ and $k_{x}=g$. Here, the mass of the particle and Planck's constant are taken to be unity.\\ 

In this case, the modulus of the inner product $|I|$ and position space distinguishability $M$ are respectively calculated as 

\begin{equation}
\label{for}
|I|=\exp{\left[-\frac{g^{2} t^{2}}{8 \sigma^{2}_{0}}-2  g^{2}  \sigma^{2}_{0}\right]} 
\end{equation}
 and 
\begin{equation}
\label{opr}
 M=\exp{\left[-\frac{g^{2} t^{2}}{8 \sigma^{2}_{t}}\right]}
\end{equation}
where $\sigma^{2}_{t}=\sigma^{2}_{0}(1+  t^{4}/4  \sigma^{4}_{0})$ .

Now, for specific choices of relevant parameters $g$, $\sigma_0$ and $t$, it can easily be shown that $M > I$.\\

 If we take squeezed state \cite{col,robi} as the initial pointer  is of the form 
\begin{equation}
\psi_{0}(\textbf{x})=\frac{1}{(2 \pi (\sigma_{0}(1+i C))^{2})^{1/4}}\exp{\left[-\frac{{x}^{2}}{4 \sigma_{0}^{2}(1+i C)^{2}}\right]}
\end{equation}
 the  post-interaction pointer states are given by 
\begin{equation}
\psi_{\pm}(\textbf{x},t)=\frac{1}{(2 \pi s'^{2}_{t})^{1/4} }\exp{[-\frac{(x \mp \frac{g t}{2})^{2}}{4 \sigma_{0} s'_{t}} \pm i k_{x} x]}
\end{equation}
where $s'_{t}=\sigma_{0}(1+iC)(1+\frac{i t}{2 m (\sigma_{0}(1+iC))^{2}})$ and $C$ is a constant.\\ 

 In this case, the mutual orthogonality between post-interaction states and position space distinguishability  are respectively 
\begin{equation}
|I|=\exp{\left[-\frac{g^{2} t^{2} }{8 \sigma^{2}_{0}}-2  g^{2} \sigma^{2}_{0}\left(1+C^{2}+\frac{ C\  t}{ \sigma^{2}_{0}}\right)\right]} 
\end{equation}
and
\begin{equation}
M=\exp{\left[-\frac{g^{2} t^{2}}{8 \sigma^{2}_{0}\left(1+(C+\frac{ t}{2 \sigma^{2}_{0}})^{2}\right)}\right]}
\end{equation}
Now, for the following choices of parameters, $\textit{t}=0.0001$, $g=10$, $C=-100$ and $\sigma=10^{-4}$, one obtains 
$ M\approx1$ and $|I|\approx 0$. 
   
For the above two examples of initial pointer states, it is shown that although the post-interaction pointer states are separated in complex space but they can be indistinguishable in real position space. Thus, the Gaussian and the squeezed states cannot be served as a faithful pointer for von Neumann measurement. In the following, we shall derive a family of faithful pointer states for which the degree of formal and operational (non)idealness will be the same. In other words, for such a family of pointer states $M=|I|$ is always satisfied. 

\section{In search of a faithful pointer for von Neumann measurement}
In the previous section we have shown why Gaussian and squeezed pointer state cannot be treated as faithful pointers.  Here we derive a class of faithful pointers and at the end of this section we provide the verification in support of our claim. In order to derive such a class of   pointers, we use the standard procedure of variational calculus. For this, let us consider that the pointer initial state $\psi(r)$  is of the form
\begin{equation}
\psi(r)=\psi_{0}(r)+\epsilon \ \ \eta(r)  
\end{equation}
 We assume that the wave function $\psi_{0}(r)$ represents a family of faithful pointer states we wish to derive. The quantity $\eta(r)$ is perturbation function with the boundary condition $\eta(\infty)=\eta(-\infty)=0$ and $\epsilon$ is complex number constant called perturbation constant.

The pointer states after interaction with the Hamiltonian  is  $\psi(r \pm s)$ where $s$ is considered to be integer and it  is used for the shift of the initial pointer corresponding to the eigenvalues of the observable $\hat{\sigma_x}$. In comparison to the earlier discussion, $|\psi(r \pm s)\rangle \equiv |\psi_{\pm}(x)\rangle$. Then, the real space distinguishability and mutually orthogonality between post-measurement states can be re-written as
\begin{equation}
M=\int_{-\infty}^{\infty} |\psi(r+s)| |\psi(r-s)| dr
\end{equation}
and
\begin{equation}
 I = \int_{-\infty}^{\infty}  \psi^{*}(r+s) \psi(r-s)dr = \langle\psi(r + s)|\psi(r - s)\rangle  \equiv R[\psi] e^{i\theta}
\end{equation}
where  $R[\psi]=\int_{-\infty}^{\infty}  \psi(r+s) \psi(r-s)dr$ is the scaler product between the post interaction pointer states.\\

Let us now invoke the commonly used Lagrange undetermined multiplier method for the optimization of a problem with suitable constraints. In order to achieve our goal of finding faithful pointer we optimize $\psi_{0}(r)$ with the constraint that $\psi(r)$ is normalized to unity and $M=|I|$. Then, the Lagrangian $(\mathcal{L})$ with the given constraints can be written as 
\begin{eqnarray}
\label{lag}
\mathcal{L}&=&\int_{-\infty}^{\infty} \psi^{*}(r+s) \psi(r-s)dr \int_{-\infty}^{\infty} \psi(r'+s) \psi^{*}(r'-s)dr'\\
\nonumber
&-&\int_{-\infty}^{\infty} |\psi(r+s)| |\psi(r-s)|dr  \int_{-\infty}^{\infty} |\psi(r+s)| |\psi(r-s)|dr \\
\nonumber
&+& \lambda \left[ \int_{-\infty}^{\infty} |\psi(r)|^{2}dr-1 \right]  
\end{eqnarray}

The quantity $\mathcal{L}$ can be maximized with respect to $\psi_{0}(r)$ with the conditions $\frac{\partial \mathcal{L}}{\partial \epsilon}|_{\epsilon=0}=\frac{\partial \mathcal{L}}{\partial \epsilon^{*}}|_{\epsilon^{*}=0}=0$. We can derive  $\frac{\partial \mathcal{L}}{\partial \epsilon^{*}}|_{\epsilon^{*}=0}$ by using Eq.(\ref{lag}) is given by
\begin{eqnarray}
\frac{\partial \mathcal{L}}{\partial \epsilon^{*}}|_{\epsilon^{*}=0}&=&R[\psi] e^{2i\theta}\int_{-\infty}^{\infty}\psi_{0}(t_{1}+2s) \eta(t_{1}) dt_{1}\\
\nonumber
&+& R[\psi] e^{- 2 i \theta} \int_{-\infty}^{\infty} \psi_{0}(t_{2}-2s) \eta(t_{2})dt_{2}+\lambda\int_{-\infty}^{\infty}\psi_{0}(r) \eta^{*}(r)dr\\
\nonumber
&-& M \left[\int_{-\infty}^{\infty} e^{- 2 i \theta}[\psi_{0}(t_{1}+2s)\eta^{*}(t_{1})] dt_{1}+  \int_{-\infty}^{\infty} e^{2 i \theta} [\psi_{0}(t_{2}-2s)\eta^{*}(t_{2})] dt_{2}\right]
\end{eqnarray}

where $r-s=t_1$ and $r'+ s=t_2$.\\
For maximization of $\mathcal{L}$ using $\frac{\partial \mathcal{L}}{\partial \epsilon}^{*}|_{\epsilon^{*}=0}=0$, we obtain the following condition to be satisfied by $\psi_{0}(r)$ is given by

\begin{eqnarray}
\lambda\psi_{0}(r)&=&-[ R[\psi]- Me^{-4 i \theta}]e^{2 i \theta}\psi_{0}(r+2s)\\
\nonumber
&-&[ R[\psi]- M e^{4 i \theta}]e^{-2 i \theta}\psi_{0}(r-2s)
\end{eqnarray}
which can be rearranged as
\begin{eqnarray}
\label{s}
\psi_{0}(r)=-\gamma_{1}e^{2 i \theta}\psi_{0}(r+2s)-\gamma_{2}e^{-2 i \theta}\psi_{0}(r-2s)]
\end{eqnarray}
where 
\begin{eqnarray}
\gamma_{1}=\frac{  R[\psi]- Me^{-4 i \theta}}{\lambda}   
\end{eqnarray}
and $\gamma_{2}=\gamma^{*}_{1}$. It is seen from Eq.(\ref{s}) that it can produce a sequence \cite{silva} of the form 
\begin{eqnarray}
\psi_{m}(r)=-\gamma_{1} e^{2 i \theta}\psi_{m+1}(r)+-\gamma_{2}e^{-2 i \theta} \psi_{m-1}(r)
\end{eqnarray}

where $m$ is an integer. If $\psi_{m}(r)$ is post measurement pointer state  then $\psi_{m+1}(r)$ and $\psi_{m-1}(r)$ are the post measurement pointer shift in upward and downward direction respectively. Since any valid wave-function needs to converge at infinity we check the convergence of $\psi_{m}(r)$. The $\psi_{m}(r)$ converges to zero, i.e., $\lim_{m \rightarrow {\infty}}(\psi_{m}(r))=0$ , if and only if,

\begin{eqnarray}
\label{se}
\psi_{m}(r)=\psi_{0}(r)e^{2 m u-2 i m \theta}  
\end{eqnarray}

with $u=\frac{1}{2}\cosh^{-1}\left[\frac{-1}{\gamma_{1}+\gamma_{2}}\right]$. Substituting $u$ in Eq.(\ref{se}) we obtain the pointer state as

\begin{equation}
\label{faith}
\psi_{0}(r)=\exp{\left[- m \ \ \Re \left[ \cosh^{-1}\Big(\frac{-1}{\gamma_{1}+\gamma_{2}}\Big)\right]+2 i m \theta\right]} \psi_{m}(r) 
\end{equation}
where $\Re$ denotes the real part. The function $\psi_{0}(r)$  is our desired form of the family of the pointers which is expected to be a faithful one, i.e., for $\psi_{0}(r)$ the formal and operational (non) idealness should be the same. 

Let us now verify if $\psi_{0}(r)$ is faithful one, for which the condition $M=|I|$ needs to be satisfied. For this, we explicitly consider the post-interaction pointer states by using Eq.(\ref{faith}) are the following,

\begin{equation}
\psi_{0}(r\pm s)=\psi_{0}(r)\exp{\left[ (r \pm s) \ \ \Re \left[\cosh^{-1}\Big(\frac{-1}{\gamma_{1}+\gamma_{2}}\Big)\right]-2 i (r \pm s) \theta\right]}
\end{equation}
 By using Eq.(\ref{opr}), the real space overlap can be calculated as
\begin{eqnarray}
\label{op}
M=\int_{-\infty}^{\infty} |\psi_{0}(r)|^2 \exp{\left[  2 r \ \ \Re \left[\cosh^{-1}\Big(\frac{-1}{\gamma_{1}+\gamma_{2}}\Big)\right]\right]}  dr
\end{eqnarray}
and by using Eq.(\ref{for}) the inner product of post-interaction states is given by
\begin{eqnarray}
\label{fo}
I = \exp{[4 i  s \theta]} \int_{-\infty}^{\infty} |\psi_{0}(r)|^2 \exp{\left[  2 r \ \ \Re \left[ \cosh^{-1}\Big(\frac{-1}{\gamma_{1}+\gamma_{2}}\Big)\right]\right]}  dr
\end{eqnarray}

It can be seen that for $\psi_{0}(r)$ the desired condition $M=|I|$ is satisfied. Hence, $\psi_{0}(r)$ given by Eq.(\ref{faith}) is a family of pointers which can be considered as a faithful one. The derived class of faithful pointers is mathematical and how such a pointer can be physically manifested is an interesting question. In a separate work, we plan to provide the details about how such a class of pointers can be produced experimentally.  

\section{Summary and conclusions}
We pointed out that in the context of 'ideal' von Neumann projective measurement for qubit system, the mutual orthogonality (formal idealness) between the post-interaction pointer states does \emph{not} ensure the macroscopic separation (operational idealness) between the probability  distributions in real space corresponding to the post-interaction pointer states. While the latter implies the former but the converse does not hold good. However, the real space distinguishability plays a crucial role for verifying the experimentally probabilities of different outcomes. It is shown that for commonly referred Gaussian class of pointers the measurement scenario can be made fully nonideal operationally  even if it is formally ideal. Such a class of pointers are then not useful if one sticks to the von Neumann projective measurement. We posed the question whether there is a particular class of pointers available (which we termed as faithful pointer) for which the degree of operational and formal (non)idealness will be the same.  In this paper, we derived a class of faithful pointers for which the  formal idealness guarantees the operational idealness. In other words, for such pointers whenever the perfect entanglement between the system and apparatus states is established, the corresponding post-interaction pointer distributions are distinguishable in real space. In that case no operational quantity (such as, error measure) is needed to be introduced. Further studies are called for to gain insights regarding the physical manifestation and  usefulness of faithful pointers in quantum measurement theory. 

\section*{Acknowledgments}
AKP acknowledges the support from Ramanujan Fellowship research grant(SB/S2/RJN-083/2014).

\end{document}